**Tuning the charge transfer in $F_x$-TCNQ/rubrene single-crystal interfaces**

*Yulia Krupskaya\*, Ignacio Gutiérrez Lezama, Alberto F. Morpurgo\**

Dr. Y. Krupskaya, Dr. I. Gutiérrez Lezama, Prof. A. F. Morpurgo
Department of Quantum Matter Physics (DQMP) and Group of Applied Physics (GAP),
University of Geneva, 24 quai Ernest-Ansermet, CH - 1211 Geneva, Switzerland
E-mail: y.krupskaya@ifw-dresden.de, alberto.morpurgo@unige.ch

Interfaces formed by two different organic semiconductors often exhibit a large conductivity, originating from transfer of charge between the constituent materials. The precise mechanisms driving charge transfer and determining its magnitude remain vastly unexplored, and are not understood microscopically. To start addressing this issue, we have performed a systematic study of highly reproducible single-crystal interfaces based on rubrene and $F_x$-TCNQ, a family of molecules whose electron affinity can be tuned by increasing the fluorine content. The combined analysis of transport and scanning Kelvin probe measurements reveals that the interfacial charge carrier density, resistivity, and activation energy correlate with the electron affinity of $F_x$-TCNQ crystals, with a higher affinity resulting in larger charge transfer. Although the transport properties can be described consistently and quantitatively using a mobility-edge model, we find that a quantitative analysis of charge transfer in terms of single-particle band diagrams reveals a discrepancy ~100 meV in the interfacial energy level alignment. We attribute the discrepancy to phenomena known to affect the energetics of organic semiconductors, which are neglected by a single-particle description –such as molecular relaxation and band-gap renormalization due to screening. The systematic behavior of the $F_x$-TCNQ/rubrene interfaces opens the possibility to investigate these phenomena experimentally, under controlled conditions.

## 1. Introduction

Organic semiconductors based on small conjugated molecules typically possess large band gaps and, in their undoped form, are essentially insulators. Rather unexpectedly, interfaces between two such materials often exhibit an electrical conductance exceeding by several orders of magnitude that of each of the two constituent materials taken separately.[1-10] This large excess conductance originates from charge transferred from one material to the other, confined within the first few molecular layers near the interface. The amount of interfacial charge is expected to depend on the electron affinity and ionization potential of the two materials or, equivalently, on the difference in their chemical potentials.[11, 12] Since overall charge neutrality needs to be preserved, it is also expected that charge transfer results in the same density of electrons and of holes at the interface. However, virtually no work has been done to test experimentally the validity of these expectations and very little is known about the microscopic mechanisms responsible for interfacial charge transfer between organic materials.

Past experimental work has probed the electronic properties of organic interfaces using both spectroscopic techniques[13-17] and transport measurements.[1-9] Significant interfacial charge transfer has been detected in virtually all cases, but no trend common to all organic interfaces has been identified. This is so even if we confine our attention exclusively to organic interfaces based on single crystals,[3, 5-9] which typically offer the most controlled experimental conditions for the investigation of interfacial charge transfer. Recent work on Schottky-gated rubrene/PDIF-CN$_2$ single-crystal heterostructures,[6, 10] for instance, has succeeded in describing interfacial charge transfer quantitatively in terms of single-particle band diagrams, i.e., a description analogous to the one used for heterostructures of conventional inorganic semiconductors.[11] This approach, however, cannot realistically account for transfer of charge much in excess of $5 \cdot 10^{12}$-$10^{13}$ cm$^{-2}$, and it cannot therefore explain the behavior of TTF/TCNQ single-crystal interfaces,[3] where interfacial densities close to $5 \cdot 10^{14}$ cm$^{-2}$ have been estimated from conductivity measurements. Additionally, even though the presence of both electrons and holes is expected from basic charge neutrality considerations, in no case both types of charge carriers have been simultaneously detected experimentally. In rubrene/PDIF-CN$_2$ interfaces, for instance, gate-dependent conductivity and Hall effect measurements have shown unambiguously that transport is dominated by electrons at the surface of PDIF-CN$_2$.[6] In all other single-crystal interfaces that have been probed, the type (or types) of charge carriers contributing to transport has not been identified experimentally.

It seems clear that reaching a proper understanding of charge transfer at organic interfaces still requires a very considerable amount of work. An effective way to proceed is to perform investigations enabling the comparison of interfaces between organic semiconductors in which –ideally– only one experimental parameter is varied controllably and systematically. To start research in this direction, here we report comparative experiments on a class of high-quality organic single-crystal interfaces, formed by rubrene (as donor) and different members of the F$_x$-TCNQ family (as acceptors).[18] Since an increase in the number of fluorine atoms lowers the energy of the molecular levels in the F$_x$-TCNQ molecules,[19] comparing the behavior of these interfaces enables a systematic analysis of charge transfer upon increasing electron affinity. Through resistivity and Hall effect measurements we identify the type of carriers that mediate interfacial transport, and determine their density and mobility. The measurements show that interfacial charge transfer does indeed increase upon increasing the electron affinity of the acceptor material. They also allow us to develop a simple physical scenario that correctly captures the observed transport properties. With this scenario in mind, we perform scanning Kelvin probe measurements to analyze the relative alignment of the valence and conduction bands for the three different F$_x$-TCNQ/rubrene interfaces. The experimental results exhibit a very systematic behavior, which –at a quantitative level, with a precision on the ~100 meV scale– is not compatible with an explanation of the observed charge transfer in terms of single-particle band diagrams. Our findings imply that a precise understanding of the energetics of organic interfaces requires taking into account physical phenomena specific to organic semiconductors, for which we discuss several possible examples.

## 2. Results and Discussion
### 2.1. Interfacial transport properties and their interpretation

All interfaces investigated in this work were realized using single crystals of $F_4$-TCNQ, $F_2$-TCNQ, TCNQ and rubrene, grown by physical vapor transport.[20] As a characterization step, we first investigated transport properties of the crystals (identical to those used to assemble the interfaces) via field-effect transistor (FET) devices, in so-called air gap stamp configuration with vacuum or air acting as gate dielectric.[21] Through this characterization, we found that holes in our rubrene devices typically exhibit a room temperature mobility between 12 and 20 $cm^2V^{-1}s^{-1}$ (in agreement with previous studies).[22-24] For electrons at the surface of $F_2$-TCNQ crystals the room-temperature mobility is approximately 6-7 $cm^2V^{-1}s^{-1}$, and increases up to ~20 $cm^2V^{-1}s^{-1}$ upon lowering temperature down to 150 K; in TCNQ and $F_4$-TCNQ devices, in contrast, electrons exhibit a much lower mobility of 0.1-0.2 $cm^2V^{-1}s^{-1}$, with a thermally activated temperature dependence.[18] The assembly of $F_x$-TCNQ/rubrene interfaces was done by means of manual lamination in air, i.e., by using a process similar to that employed in the realization of air-gap stamp single-crystal FETs.[25] (see experimental section for details). The optical microscope image in **Figure 1**a illustrates an example of a representative device.

Figure 1b shows representative room-temperature, multi-terminal *I-V* characteristics measured on the three studied interfaces, $F_4$-TCNQ/rubrene, $F_2$-TCNQ/rubrene and TCNQ/rubrene. We measured (approximately five) different devices for each kind of interface, all exhibiting an excellent level of device-to-device reproducibility. In all cases, the *I-V* characteristics show linear behavior without any noticeable contact effects. Moreover, the comparison of two- and four-terminal measurements (which show comparable resistivity values) indicates that the resistance of the channel is much larger than the contact resistance and therefore the contact resistance can be neglected. Since the room-temperature resistance measured on individual crystals of each molecule separately is much larger than $10^8$ Ω, the resistance values found when measuring the interfaces are indicative of a strongly enhanced interfacial conductivity (i.e., of interfacial charge transfer). Specifically, the room-temperature resistivity $\rho$ of the three different interfaces was measured to range (in different devices) between 390-590 kΩ/□ for $F_4$-TCNQ/rubrene, 1-1.5 MΩ/□ for $F_2$-TCNQ/rubrene, and 3.5-7 MΩ/□ for TCNQ/rubrene. The resistivity values for $F_4$-TCNQ/rubrene are the lowest among all organic charge transfer interfaces reported so far, except for TTF/TCNQ.[3]

To establish whether the carriers mediating transport are electrons or holes (or both), and to determine their density and mobility, we have measured the Hall effect. Figure 1c shows that the Hall resistance increases linearly with increasing magnetic field with a positive slope for all the three interfaces, indicating that holes in rubrene dominate the observed transport properties in all cases. Note that this observation –by itself– does not exclude the possibility that electrons in $F_x$-TCNQ may also contribute to transport. Indeed, in the presence of a comparable density of holes and electrons, the sign of the Hall resistance is determined by the carriers with the highest mobility, so that –strictly speaking– the Hall effect data shown in Figure 1c simply indicate that holes in rubrene have higher mobility than electrons in $F_x$-TCNQ. Nevertheless, if we do assume that the electron contribution to transport is negligible (i.e., the mobility of electrons is much lower than that of holes), we can extract the hole density $n_h$ from the Hall effect and the hole mobility from the measured resistivity. We find $n_h$ = $1\cdot10^{13}$, $2.6\cdot10^{12}$ and $1.3\cdot10^{12}$ cm$^{-2}$ for F$_4$-TCNQ/rubrene, F$_2$-TCNQ/rubrene and TCNQ/rubrene, respectively; all three interfaces exhibit essentially the same value of hole mobility, $\mu$ = 1.3-1.6 cm$^2$V$^{-1}$s$^{-1}$. This very systematic trend in both $n_h$ and $\mu$ is totally consistent with transport due to holes in rubrene only and does not appear to be compatible with a significant electron contribution (if the contribution of electrons to transport could not be neglected, the behavior of interfaces based on F$_2$-TCNQ –where the electron mobility is very high– should not fit a trend consistent with that of interfaces based on TCNQ and F$_4$-TCNQ). On this experimental basis, we conclude that for interfaces based on the three different F$_x$-TCNQ crystals, holes at the surface of rubrene determine the transport properties and electrons can be considered as fully localized (the precise microscopic mechanism responsible for their localization remains to be determined).

To verify the consistency of this conclusion, we analyze the transport data in terms of a simple physical scenario that has been developed to describe organic single-crystal transistors, and that allows us to rationalize our observations in terms of a mobility edge model.[26, 27] In particular, the presence of large density of localized electrons at the surface of the F$_x$-TCNQ crystals causes strong electrostatic potential fluctuations at the surface of rubrene, resulting in two main effects. The first is the presence of considerable disorder in rubrene that broadens the top of the valence band and results in a tail of localized states (as illustrated in the inset of **Figure 2**a). The second is a suppression of the mobility of holes, consistent with experimental observations: indeed the mobility of holes at F$_x$-TCNQ/rubrene interfaces is only 1.5 cm$^2$V$^{-1}$s$^{-1}$, much smaller than the mobility of holes in rubrene FETs with vacuum gate dielectric (15-20 cm$^2$V$^{-1}$s$^{-1}$)[22] and also of the mobility measured in rubrene FETs with polymeric gate dielectric[23, 24, 28] having dielectric constant comparable to F$_x$-TCNQ crystals (~ 5-10). In this scenario the chemical potential in rubrene is located in the band tail, so that the density of holes $n_h$ mediating transport decreases exponentially with lowering temperature, with an activation energy $E_a$ determined by the distance between the chemical potential and the position of the mobility edge. A simple calculation assuming Boltzmann statistics for the thermally excited holes gives:[26, 5]

$$n_h = \int_{E_a}^{\infty} \frac{N}{W} \exp(-\frac{E}{k_B T}) dE = \frac{N}{W} k_B T \exp\left(-\frac{E_a}{k_B T}\right). \qquad (1)$$

where $N$ is the number of molecules per unit area at the surface of a rubrene crystal forming the interface with a F$_x$-TCNQ crystal ($N \sim 1.2\cdot10^{14}$ cm$^{-2}$, estimated from the crystals structure); $W$ is the width of the valence band in rubrene ($W \sim 0.4$ eV);[29, 30] $N/W$ therefore corresponds to the density of states in the rubrene valence band. **Equation 1** is a relation between the density of holes $n_h$ involved in transport, and their activation energy $E_a$. This relation can be checked experimentally to verify the validity of the scenario that we have proposed.

To this end, we determine $E_a$ from temperature-dependent resistivity measurements (Figure 2a), which exhibit an almost perfect exponential behavior (Figure 2b; the exponential behavior of $\rho$ can be attributed to $n_h$ because the temperature dependence of the intrinsic mobility in organic semiconductors is much weaker than exponential). We find $E_a$ = 15, 30 and 45 meV for F$_4$-TCNQ/rubrene, F$_2$-TCNQ/rubrene and TCNQ/rubrene interfaces, respectively (in all cases the error is approximately 5 meV). Inserting these values into Equation 1, we obtain the (room-temperature) values for $n_h$ summarized in **Table 1**. The agreement with the hole density extracted from Hall effect measurements is surprisingly good for both TCNQ/rubrene and F$_2$-TCNQ/rubrene (better than 10%). Given the simplicity of the model, it is good also for F$_4$-TCNQ/rubrene, in which the larger discrepancy –a factor of 2– likely originates from the fact that for F$_4$-TCNQ $E_a$ is comparable to $k_B T$ for a part of the temperature interval in which the measurements are performed (i.e., at high temperature the behavior is not truly exponential). We conclude that a rather simple scenario captures the basic aspects of interfacial transport as due to holes at the surface of rubrene crystals. Moreover, all the observed trend –in the density of holes as well as in their activation energy– evolve systematically with increasing the amount of fluorine atoms in the F$_x$-TCNQ molecules, i.e. with increasing the electron affinity of the acceptor material.

### 2.2. Energetics of F$_x$-TCNQ/rubrene interfaces

Having gained, thanks to the systematic trend found in the experiments, a good physical understanding of the transport measurements in the F$_x$-TCNQ/rubrene family of interfaces, we proceed with the analysis of their energetics. To do this, it is necessary to determine the relative alignment of the bands, as well as positions of the chemical potentials, for all the materials considered.

A reliable estimate of the distance between the chemical potential and the relevant transport band in high-quality organic semiconductors is provided by the activation energy of the conductance measured in single-crystals of the individual materials, which is straightforward to determine experimentally. Once the position of the chemical potential with respect to the band is known, scanning Kelvin probe measurements across the interface between two crystals can be used to extract the relative position of the chemical potentials, and from that, the relative band alignment. The distance between the chemical potential and the top of the valence band in rubrene single-crystals has been determined in the past using different techniques and it is known to be approximately 300 meV.[31, 32, 6] We have measured the activation energy of electron transport in individual F$_4$-TCNQ, F$_2$-TCNQ and TCNQ single-crystals and found in all cases values close to 200 meV. The difference $\Delta E_F$ between the chemical potentials of rubrene and F$_x$-TCNQ crystals, which corresponds to the contact potential difference, was measured using scanning Kelvin probe microscopy (SKPM)[33] on the same heterostructures employed to perform the transport measurements. As an example, representative atomic force microscopy (AFM) and SKPM images for F$_4$-TCNQ/rubrene are presented in **Figure 3**a and 3b (see the experimental section for details). Extracting the contact potential difference by scanning across the interfaces (rather than measuring individual rubrene and F$_x$-TCNQ crystals separately) is particularly effective, because it enables the contact potential to be measured directly, independently of the work function of the tip. Line-scan topography and contact potential measurements for all F$_x$-TCNQ/rubrene interfaces are shown in Figure 3c-e and 3f-h. By looking at the height of the potential step we find that $\Delta E_F \sim$ 250, 200 and 100 meV, for $F_4$-TCNQ/rubrene, $F_2$-TCNQ/rubrene and TCNQ/rubrene, respectively.

**Figure 4** summarizes all information about the relative position of the bands in rubrene and the different $F_x$-TCNQ crystals, as well as the position of $E_F$, extracted from the measurements. The diagrams represent the band alignment prior to charge transfer. It is apparent that –at a qualitative level– they exhibit a systematic behavior consistent with the results of the transport measurements. Specifically, the difference in the values of $E_F$ between acceptor ($F_x$-TCNQ) and donor (rubrene) materials increases with increasing the number of fluorine atoms, in agreement with expectations. As a result, a larger electron affinity leads to a larger amount of charge transfer (compare Figure 4 with the values of hole density reported in Table 1). Despite this remarkably systematic behavior of the experimental results, the observed transport properties cannot be reproduced at a precise quantitative level in terms of these band diagrams, as we discuss below.

It is obvious (from basic thermodynamics) that the position of the chemical potential at the interface after charge transfer necessarily should lie in between the initial $E_F$ values of rubrene and $F_x$-TCNQ crystals. In Figure 4 we observe that, even if we neglect the effect of the interfacial electrical dipole (which is generated by the transferred charge and pushes the valence band of rubrene and conduction band of $F_x$-TCNQ further apart, acting against the charge transfer),[6] the chemical potential at the interface can at most lie as low as the initial value of $E_F$ in $F_x$-TCNQ crystal. Within a rigid band model, this implies that the distance between $E_F$ at the interface and the valence band of rubrene (or, with the same terms used to describe transport, the mobility edge in rubrene) cannot be smaller than 50, 100 and 200 meV for $F_4$-TCNQ, $F_2$-TCNQ and TCNQ, respectively. These values are in all cases larger than the activation energy of the conductance in the three different interfaces (15, 30, and 45 meV; see Table 1). If the effect of the interfacial electrical dipole was included, the quantitative mismatch would be even more significant. We conclude that, despite the systematic behavior of the $F_x$-TCNQ/rubrene interfaces and the qualitative trends all in agreement with our physical expectations, a simple rigid band model –i.e., the same type of model commonly used in conventional inorganic semiconductors–[11] is not sufficiently accurate at a quantitative level on the energy scale probed by transport.

Finding that –to be accurate on an energy scale of ~100 meV– a quantitative description of $F_x$-TCNQ/rubrene interfaces has to go beyond the simplest single-particle rigid band model, normally used for conventional inorganic semiconductors, is not surprising. On such an energy scale, physical phenomena that are irrelevant for inorganic semiconductors can certainly contribute to the energetic balance in organic materials. In most organic molecules, for example, the reorganization energy –i.e., the energy associated to the change in the molecular structure when passing from its neutral to its ionized form– is typically of the order of 100 to 200 meV.[34, 29, 35] When, due to charge transfer, two neutral rubrene and $F_x$-TCNQ molecules on opposite sides of the interface become respectively positively and negatively charged, the corresponding energy gain should be considered. The relevance of this contribution is likely to depend on how delocalized the charges are in the organic materials:

the more delocalized the less the reorganization energy will be relevant (which is why in high-quality inorganic semiconductors this phenomenon is irrelevant). In very simple terms, this is because if charges are delocalized, the time that they spend on each molecule is too short to cause the molecular structure to relax. At a quantitative level, therefore, this process may be more relevant in interfaces where transport is hopping-like (i.e. thermally activated conductivity) as compared to interfaces exhibiting band-like transport. This may be the reason why in PDIF-CN$_2$/Ruberene interfaces, in which electron transport is band-like, a rigid band model was found to work quantitatively, on an energy scale of few tens of meV.[6]

Another process that is known to have large effects in organic materials is band-gap renormalization due to electrostatic screening.[36, 37] To understand the phenomenon, we recall that the band gap of an organic semiconductor corresponds –in a first approximation– to the HOMO (highest occupied molecular orbital)-LUMO (lowest unoccupied molecular orbital) gap of the constituent molecules. In simple terms, the HOMO-LUMO gap originates from size quantization determining the single-particle level spacing in the molecules, and from Coulomb interactions between the electrons. This last contribution can be very significantly suppressed by changing the electrostatic environment around the molecules (or in other terms, by electrostatic screening), as it can be checked by comparing the band gap in bulk materials and in monolayers (or thin films) deposited on metal surfaces. For instance, the HOMO-LUMO gap measured for a monolayer of C$_{60}$ molecules deposited on a metallic surface was found to be of ~1 eV smaller than in a crystal.[36] Similarly, for pentacene thin films deposited on metal substrates, the HOMO-LUMO gap measured by photo-emission spectroscopy at the film surface decreases continuously with decreasing the film thickness (a dependence on thickness is visible experimentally up to ~ 10 nm).[37] This is because upon increasing the thickness the distance of the exposed surface from the metal substrate also increases, and – correspondingly– the effect of screening decreases. At organic interfaces, the occurrence of charge transfer is likely to change the screening properties, which may cause a change in the HOMO-LUMO gap on an energy scale that is relevant for a quantitative accurate description of transport. If this is so, the assumption of a rigid band model –namely, that the band gap of the materials is a given constant in the bulk and at the interface– is not valid. In F$_x$-TCNQ/rubrene interfaces a narrowing of the semiconducting gap close to the interface would result in a smaller distance between the chemical potential in rubrene and the edge of the valence band, as it is needed to achieve quantitative agreement with transport measurements.

## 3. Conclusion

In conclusion, we have performed systematic experiments on F$_x$-TCNQ/rubrene interfaces to identify the type of interfacial charge carrier present, to determine their concentration, and to investigate the relative alignment of the different energetic levels in these systems (top of valence band of rubrene; bottom of conduction band in F$_x$-TCNQ; the relative position of the chemical potential in the different materials). We find that holes at the surface of rubrene are in all cases mobile, and exhibit a room temperature mobility just above 1 cm$^2$/Vs, independent of the electron accepting materials forming the interface. Through Hall effect measurements we determine the density of holes, which we find to systematically increase upon increasing the content of fluorine atoms in the F$_x$-TCNQ crystal considered. We have analyzed the relative alignment of the chemical potential in the two materials and their distance from the respective bands and found that, although the qualitative behavior systematically agrees with expectations, at a quantitative level there is a mismatch of approximately 100 meV between the estimated position of the chemical potential relative to the valence band of rubrene and the activation energy measured in transport. This mismatch is likely to originate from specific aspects of the electronic properties of organic materials –such as the effect of the reorganization energy and the presence of a significant band-gap renormalization due to screening– which need to be considered to gain a complete understanding of the interfacial electronic structure. Future studies should take advantage of the very systematic behavior that has been demonstrated in the experiments presented here, to analyze more in detail these aspects of the electronic properties, and their effect on the interfacial electronic structure.

**4. Experimental Section**

Organic single crystals of $F_4$-TCNQ, $F_2$-TCNQ, TCNQ and rubrene (orthorhombic polymorph) were grown by physical vapor transport[20] in a furnace consisting of a quartz tube with a resistive heater used to establish a temperature gradient, in the presence of a flow of ultrapure argon gas. The starting material was placed in the hotter end of the furnace and the sublimed molecules were transferred in vapor phase to the colder end of the furnace, where the crystal growth took place. Crystals of the different materials grown in this way were selected under an optical microscope, after which the $F_x$-TCNQ/rubrene interfaces were assembled manually by laminating crystals in air (using a process similar to that employed in the realization of air-gap stamp single-crystal FETs).[25] More specifically, prior to lamination of the rubrene crystals, 20 nm Au contacts were evaporated on the flat surface of the electron-accepting crystals (i.e., $F_4$-TCNQ, $F_2$-TCNQ or TCNQ) very shortly after their growth, using shadow masks to define a multi-terminal geometry.[6] Subsequently, freshly grown rubrene crystals of suitable dimensions were laminated on top to form the interfaces. The interfacial contact plane of rubrene crystals corresponds to the high mobility plane (*ab*-crystallographic direction).[22] The contact planes of the $F_x$-TCNQ crystals are the same as used in previous FET studies.[18]

Charge transport measurements were performed in the dark using an Agilent Technology E5270B parameter analyzer. Measurements performed in air and in vacuum gave virtually identical results. The temperature dependence of the resistivity was measured upon cooling in the range between 30 and 300 K, with the sample in the vacuum chamber of a flow cryostat. Upon warming up the devices, the initial value of the measured resistance was usually recovered (i.e., the behavior of the interfaces is reversible upon cooling and warming up the devices). The Hall effect measurements were performed using a cryo-free Teslatron system from Oxford Instruments equipped with a 12 T superconducting magnet. Scanning Kelvin probe measurements were performed in air, on the same interfaces used to investigate transport, using a commercial Asylum Cypher Scanning Probe Microscope.

**Acknowledgements**

The authors would like to thank A. Ferreira for technical support. Y. K. acknowledges the financial support from the German Research Foundation (DFG) through the Research Fellowship KR 4364/1-1.

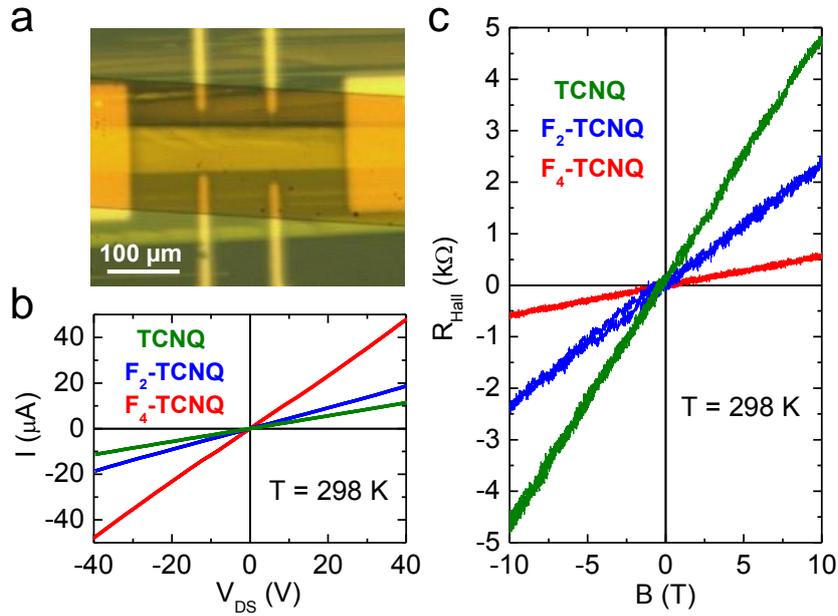

**Figure 1.** a) Optical microscope image of a representative multi-terminal $F_4$-TCNQ/rubrene device, enabling longitudinal conductivity and Hall effect measurements. A narrow rubrene crystal is laminated on top of a wider $F_4$-TCNQ crystal, after having evaporated electrical gold contacts onto $F_4$-TCNQ through a shadow mask. b) Multi-terminal *I-V* characteristics measured at room temperature on the three interfaces; the large conductance and the linearity of the *I-V* curve originate from transport at the interface. c) Hall resistance $R_{Hall}$ as a function of applied magnetic field *B*, measured at room temperature on the three interfaces. From the analysis of the Hall effect we infer that transport is dominated by holes in rubrene and extract their density and mobility.

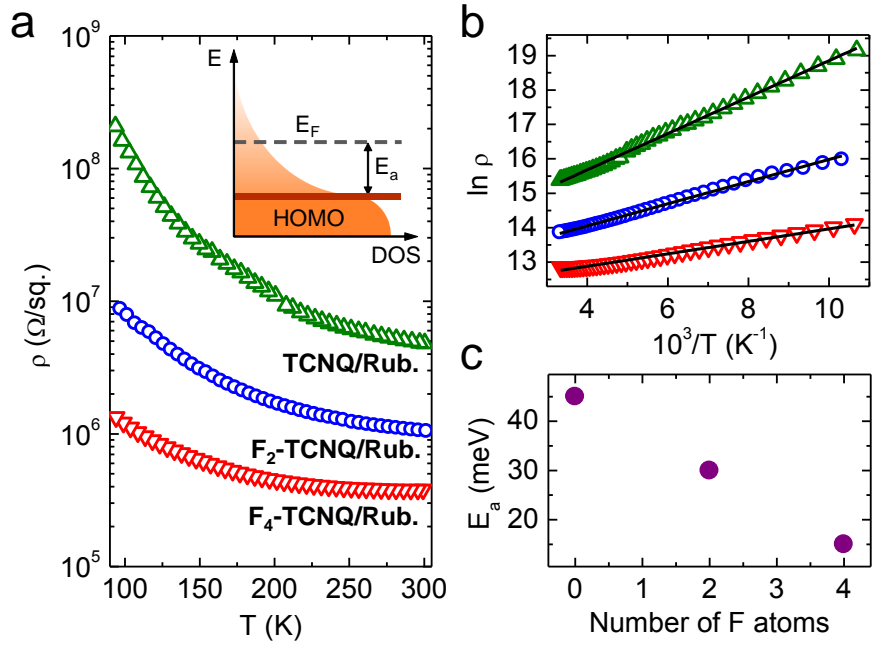

**Figure 2.** a) Temperature dependence of the resistivity for all three studied interfaces. The inset schematically represents the energy dependence of the density of states (DOS) at the surface of rubrene crystals. The DOS is associated to the presence of a disorder-induced band tail that decays rapidly away from the top of the valence (HOMO) band (corresponding to the mobility edge; see text for more details). b) Same data as in panel a), re-plotted as $\ln\rho$ vs. $T^{-1}$ (symbols): the linear dependence indicates that the behavior of transport is thermally activated. Solid black lines represent fits done to extract the activation energy $E_a$ for the different interfaces. c) The activation energies $E_a$ extracted from the data shown in panel b) decrease systematically upon increasing the number of fluorine atoms in the acceptor $F_x$-TCNQ molecules.

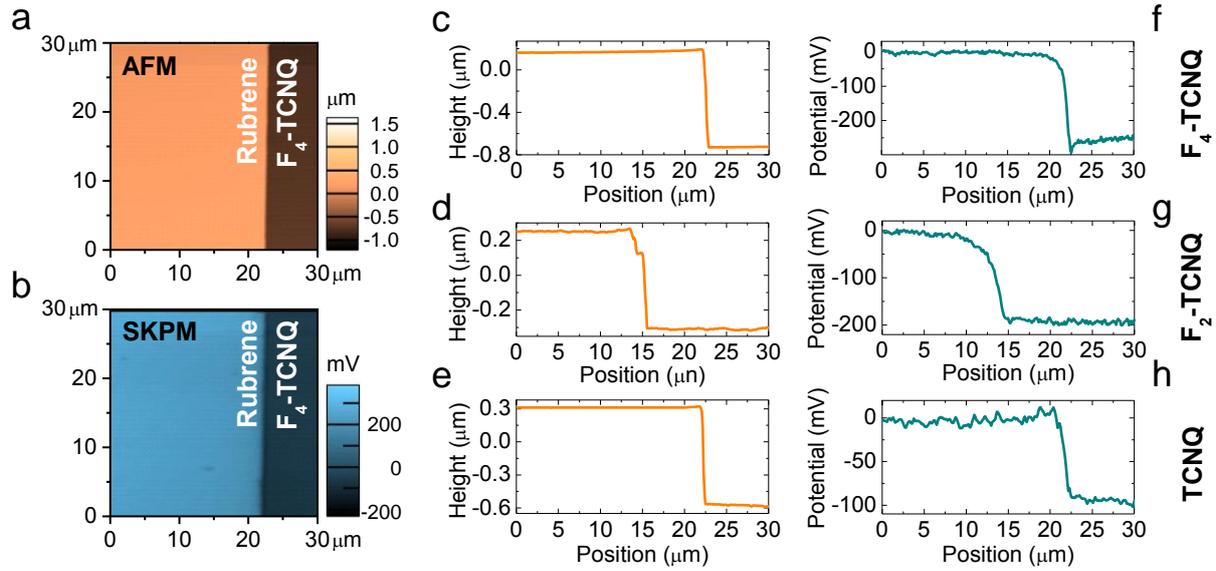

**Figure 3.** Scanning Kelvin probe microscopy (SKPM) measurements performed on the three different $F_x$-TCNQ/rubrene interfaces. AFM (a) and SKPM (b) images of a $F_4$-TCNQ/rubrene heterostructure, exhibiting a clear step in both the topography (a) and in the contact potential (b) as the tip is moved from the surface of rubrene (left side of the image) to that of $F_4$-TCNQ (right side of the image). c)-e) Line-cuts extracted from topography images of the three different interfaces, analogous to the one shown in panel a). The step height (which varies approximately between 0.5 and 1 μm) corresponds in all cases to the thickness of the rubrene crystal. f)-h) Line-cuts extracted from SKPM images of the three different interfaces, analogous to the one shown in panel b). The step corresponds to the difference in contact potential measured on the rubrene and the $F_x$-TCNQ surfaces. For ease of comparison, the contact potential of rubrene is taken as a reference and set to zero in all cases.

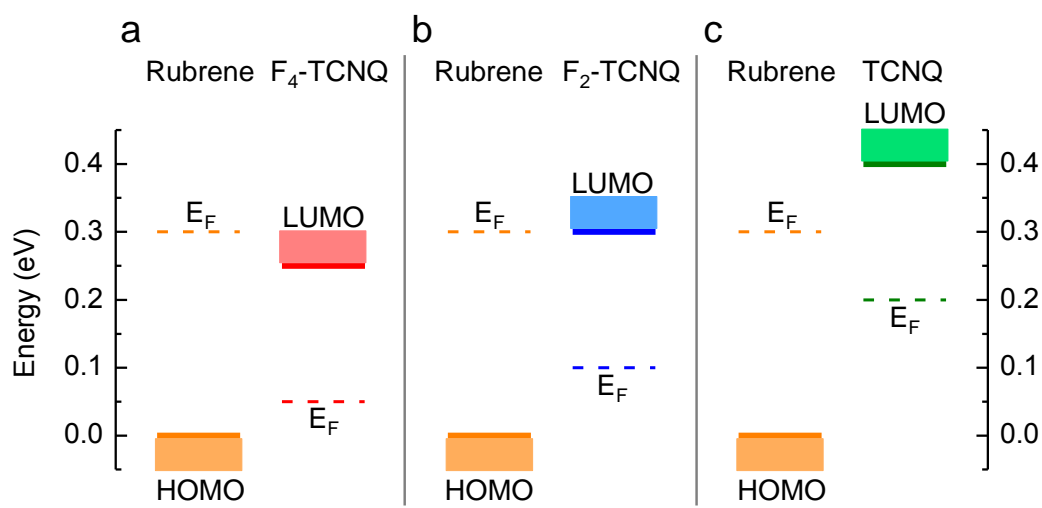

**Figure 4.** Band alignment for the individual rubrene and $F_x$-TCNQ crystals prior to charge transfer, as inferred from the analysis of SKPM and thermally activated resistivity measurements, discussed in the text. The figure summarizes all information about the relative position of the bands in rubrene and the different $F_x$-TCNQ crystals, as well as the position of $E_F$ in the different crystals (for all materials the Fermi energy is located in the disorder-induced tail of states). The difference between the chemical potential $E_F$ in rubrene and in $F_x$-TCNQ crystals increases systematically with increasing the degree of the fluorination, leading to a larger charge transfer (see text for more details).

**Table 1.** The measured activation energies ($E_a$, first column) allow us to estimate theoretically the density of interfacial charge carriers, using the mobility edge model discussed in the text. The carrier density, determined experimentally from Hall effect measurements and calculated from Equation 1, is listed in the second and third column, respectively. Note the excellent agreement between the two (in the case of F$_4$-TCNQ, the agreement is less precise because the activation energy is too small to have thermally activated transport though the entire measurement range).

| Interface | $E_a$ [meV] | $n_h$ (experiment) [cm$^{-2}$] | $n_h$ (calculated) [cm$^{-2}$] |
|---|---|---|---|
| F$_4$-TCNQ/rubrene | 15 | $1.0 \cdot 10^{13}$ | $4.3 \cdot 10^{12}$ |
| F$_2$-TCNQ/rubrene | 30 | $2.6 \cdot 10^{12}$ | $2.4 \cdot 10^{12}$ |
| TCNQ/rubrene | 45 | $1.3 \cdot 10^{12}$ | $1.4 \cdot 10^{12}$ |